\documentclass[]{llncs}
\usepackage{graphicx}
\usepackage{comment}
\usepackage{array}
\usepackage{booktabs}
\usepackage{footnote}
\usepackage{threeparttable}
\usepackage{soul}

\usepackage{geometry}
\usepackage{color}

\newcolumntype{L}[1]{>{\raggedright\let\newline\\\arraybackslash\hspace{0pt}}m{#1}}
\newcolumntype{C}[1]{>{\centering\let\newline\\\arraybackslash\hspace{0pt}}m{#1}}
\newcolumntype{R}[1]{>{\raggedleft\let\newline\\\arraybackslash\hspace{0pt}}m{#1}}

\begin{document}
\title{COVID-19 Disease Identification on Chest-CT images using CNN and VGG16}
\author{Briskline Kiruba S, Petchiammal A, D. Murugan}

\institute{Manonmaniam Sundaranar University, Tirunelveli, India\\
\email{\{kiruba.briskline,ampetchiammal\}@gmail.com,dmurugan@msuniv.ac.in}\\
}
\maketitle    
\begin{abstract}

A newly identified coronavirus disease called COVID-19 mainly affects the human respiratory system. COVID-19 is an infectious disease caused by a virus originating in Wuhan, China, in December 2019. Early diagnosis is the primary challenge of health care providers. In the earlier stage, medical organizations were dazzled because there were no proper health aids or medicine to detect a COVID-19. A new diagnostic tool RT-PCR (Reverse Transcription Polymerase Chain Reaction), was introduced. It collects swab specimens from the patient's nose or throat, where the COVID-19 virus gathers. This method has some limitations related to accuracy and testing time. Medical experts suggest an alternative approach called CT (Computed Tomography) that can quickly diagnose the infected lung areas and identify the COVID-19 in an earlier stage. Using chest CT images, computer researchers developed several deep learning models identifying the COVID-19 disease. This study presents a Convolutional Neural Network (CNN) and VGG16-based model for automated COVID-19 identification on chest CT images. The experimental results using a public dataset of 14320 CT images showed a classification accuracy of 96.34\% and 96.99\% for CNN and VGG16, respectively.


\end{abstract}
\section{Introduction}
COVID-19 is a coronavirus disease that can cause respiratory illness in humans. COVID‑19 spreads when people breathe in air contaminated by droplets and tiny airborne particles containing the virus. This largest pandemic affects more than 500 million people around the world and causes 62.7 lakhs of people deaths. In the starting year of COVID-19, RT-PCR (Reverse Transcription Polymerase Chain Reaction) laboratory test is used to identify the disease ~\cite{zhu2020novel}. This test collects swab specimens from the human nose or throat where the virus stayed. It takes 4 to 6 hours to generate results and has low sensitivity ~\cite{wang2020detection}. Apart from this, the testing kit has a severe problem and low efficiency. For all these issues, it uses the radiology method to identify the clear findings of the COVID-19 disease.

Medical experts and organizations suggest that radiology imaging shows apparent lung abnormalities and lung involvement findings. Radiology imaging-based imaging test methods are X-ray, and CT defines human chest involvement ~\cite{horry2020covid}. Among the two, X-ray has some limitations due to less sensitivity and does not show any abnormalities in the earlier stage. Due to these problems, CT(Computed Tomography) is the best diagnosis method of COVID-19, showing the evident lung abnormalities in the earlier step. With the help of radiographers, computer researchers are involved in the study of CT images implemented in deep learning models and detect lung infections efficiently.

This paper presents a deep learning-based diagnosis system for the COVID-19 virus infection assessment based on lung CT images. COVID-19 infected and normal chest CT images are collected from various public sources consisting of 14320 images. Two deep learning-based models, CNN and VGG16 are implemented to compare their performance. These models efficiently identified a COVID-19 disease and achieved an accuracy of 96.34\% and 97.40\% for CNN and VGG16, respectively.

The remaining sections of the paper is organized as follows:  Section~\ref{sec:related-work} describes the related work, followed by the dataset details in Section~\ref{sec:dataset}. Section~\ref{sec:methodology} presents the methodology and Section~\ref{sec:experimentsandresults} summarizes the results. Finally, conclusions are given in Section~\ref{sec:conclusion}.

\section{Related Work}
\label{sec:related-work}
In medical imaging, many studies related to COVID-19 identification use CT images with the help of deep learning models. Models are separated by classification, and segmentation is discussed here in a related study of COVID. In ~\cite{xu2020deep}, Residual Network (ResNet) built with various layers, looks like a pyramidal structure to classify the CT images with an accuracy of 86.70\%. Within the model of DenseNet121, they are using 757 CT image to accurately extract the infected lung areas with an accuracy result of 84.07\% ~\cite{mishra2020identifying}. 

To classify the high dimensional convolutional features with less time using the AlexNet model processed 7500 CT images also found other diseases like lung tumors to predict the disease with an accuracy of 98.25\% ~\cite{zhou2021ensemble}. EfficientNet ~\cite{silva2020covid} uniformly scales all the dimensions using compound co-efficient with 87.68\% using more than 3000 CT images. In DCNN ~\cite{wang2021deep}, using 746 CT image reduces the dimensional size and computational cost. U-net~\cite{gozes2020coronavirus} specifies the encoding and decoding transformations performed are also recommended for medical imaging segmentation to detect COVID. In this study, a combined dataset with a large number of samples is used to validate the performance of two deep learning models.

\section{Dataset}
\label{sec:dataset}
Several public COVID-19 CT image datasets are available in various data repositories like Kaggle, Git-Hub, and IEEEDataPort. Using this database, we collected 14320 chest CT images split into two classes as covid and normal, namely as All COVID Dataset - Splitted. This section describes the five  datasets combined in this work are shown in Table~\ref{tab:dataset}.

\subsection{COVID-19 Lung CT Scans} COVID-19 Lung CT Scans:This dataset consists of 746 chest CT images split into COVID and normal. Among these 349 COVID CT images from 216 studies and 347 normal CT images. Covid CT image resolution is 819x460 in size from China and is described in ~\cite{yang2020covid}.

\subsection{SARS-COV-2 CT-Scan Dataset}This dataset~\cite{soares2020sars} contains 2481 chest CT images of 1252 COVID CT images and 1229 normal with 819x460 dimensions taken from the Public Hospital of Government Employees of Sao Paulo, Brazil. 

\subsection{COVID 19 X-ray and CT Scan Image Dataset} This public dataset ~\cite{covid19x} contains 8055 CT images of 5427 COVID, and 2628 normal resolution is 512x512 in PNG format.

\subsection{COVID-19DATASET} This dataset ~\cite{chetoui2021automated}consists of a total of 1964 CT images. There are 982 images with COVID infections and the remaining 982 images from normal patients. The standard resolution of CT image size of 256x256 in png format.

\subsection{COVID-19 CT image} This dataset ~\cite{elmuogy2021efficient} consists of 1051 CT images in total. There are 501 images with COVID symptoms and the remaining 550 images from normal patients. The resolution of CT images in this dataset is 512x512 pixels in PNG format.

The final combined dataset used in this study consists of 5,785 normal and 8,535 COVID-19 infected chest CT images. This final dataset was used to validate the performance of the CNN and VGG16 models.

\begin{table}
\centering
\renewcommand{\arraystretch}{1.5}
\caption{List of COVID-19 chest CT datasets}
\label{tab:dataset}
\begin{tabular}
{lrrrr}
\toprule
\textbf{Dataset name} & {COVID} & {Normal} & {Total} & {Image Resolution} \\
\midrule 
COVID-CT-Dataset~\cite{yang2020covid} & 349 & 463 & 746 & 819 x 460 \\
SARS-COV-2 CT-Scan Dataset ~\cite{soares2020sars} & 1252 & 1229 & 2481 & 819 x 460 \\
COVID 19 X-ray and CT Scan Dataset ~\cite{covid19x} & 5427 & 2628 & 8055 & 512 x 512 \\
COVID-19DATASET ~\cite{chetoui2021automated} & 982 & 982 & 1964 & 256 x 256 \\
COVID-19 CT image ~\cite{elmuogy2021efficient} & 501 & 550 & 1051 & 512 x 512 \\ 
\midrule
Combined dataset  & 8535 & 5785	& 14320	 & 256 x 256 \\
\bottomrule
\end{tabular}
\end{table}

\section{Methodology}
\label{sec:methodology}
In this study, the proposed methodology defines automated COVID-19 detection based on lung CT images using CNN and VGG16 deep learning models. We used a large-scale combined split dataset to train the models for classifying whether the CT images are COVID or normal.

\subsection{Pre-processing}
Creating a medical image is diagnosed with the interior parts of the human body. One of the test methods is the CT imaging modality, which is used to detect COVID-19. There is a need to preprocess the CT images due to low contrast, intensity, and clarity to prevent a false diagnosis of the COVID-19 disease ~\cite{polsinelli2020light}. The reason to preprocess the image is to remove irrelevant data and increase the accuracy level performed by image resizing, segmentation, and enhancement. 

\subsection{Image Augmentation}
The image augmentation techniques are performed for image detection purposes in this dataset. Expand more features from All COVID Dataset images from different dimensions. It generally performs zoom, horizontal flip, shear, and shifting. It is mainly applied to the training dataset images; either position augmented where the pixel position of the image changes or color augmented, where the color of an image is modified by changing its pixel values. Color augmentation can be performed by varying an image's brightness, contrast, and saturation. From the above, all operations are performed to combined the images on an Combined Dataset of covid and normal. Finally, this dataset can be uploaded to Kaggle public data repository to evaluate the methodology. Using CT dataset was applied in CNN and VGG16 models to generate better accuracy.

\subsection{Convolutional Neural Network (CNN) architecture}
\begin{figure*}[t!]
\centering
\includegraphics[scale=0.5]{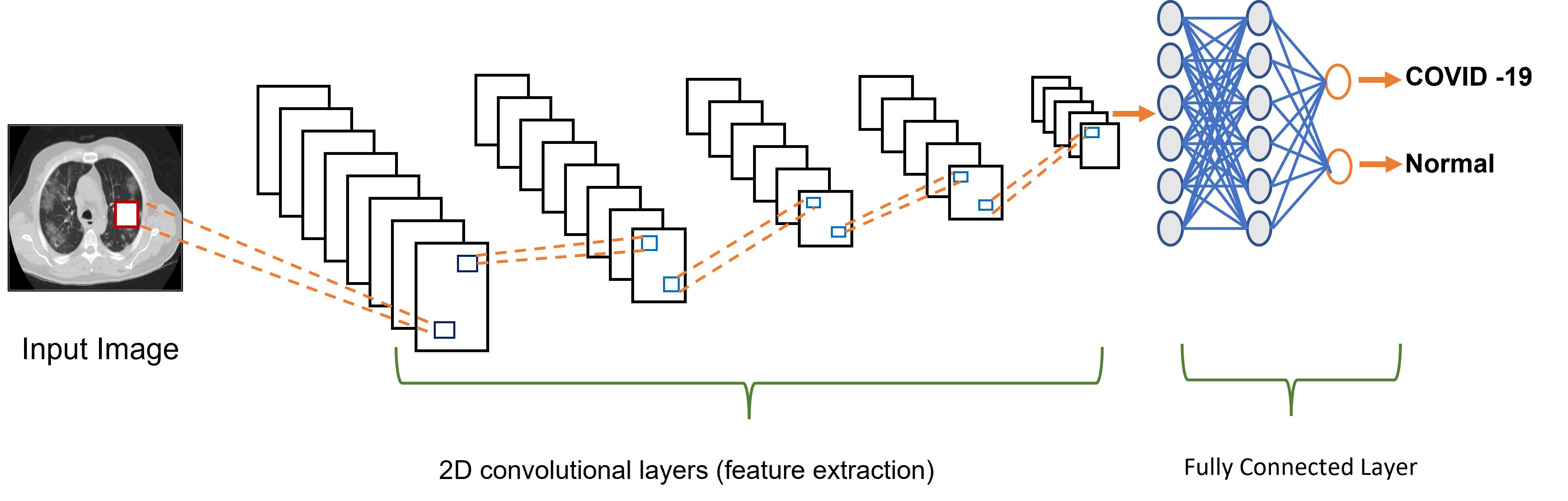}
\caption{CNN Architecture for COVID-19 Classification }
\label{fig:cnn}
\end{figure*}

A CNN has more convolution layers and handles large volumes of image data. The main work of the model is COVID CT image taken as input and assigned weights and biases for various aspects of an image and able to differentiate one another from finally extracted features from the CT image. CNN architecture has a stack of layers, namely the convolutional layer, Pooling layer, ReLU layer, and fully connected layer ~\cite{polsinelli2020light}. We proposed this model for using CT images to classify its normal or COVID, and it mainly satisfied the clinician's expectations. The CNN architecture is shown in Figure~\ref{fig:cnn}.

In this CNN architecture, the input is a CT image size of 256 x 256 fed into the convolutional and pooling layers network, which performs feature extraction. The fully connected layer acts as a classifier that performs non-linear transformations of feature extraction ~\cite{lokwani2020automated}. The CT image is an input, and it will predict whether it's a COVID or a normal CT image as an output. The layers are designed to filter size is 32, and the padding is the same with a kernel size of 3 and the activation function as ReLU. Here, the First max-pooling layer has a pool size of 2. The flat layers convert all the pooled layers into a single column. In the end, ReLU and softmax are the two dense layers formed to perform the classification as COVID or normal~\cite{kogilavani2022covid}.

\subsection{VGG 16 Architecture }
The VGG16 architecture was first developed in 2014 ~\cite{dansana2020early}. It has 16 layers where 13 convolutional layers are stacked with 3x3 filters and 2x2 max-pooling layers. A network contains three fully connected layers, and classification is performed using the softmax function ~\cite{shah2021diagnosis}, and the architecture is followed by the arrangement of convolutional and pooling layers. Finally, there are two fully connected layers, a softmax for output. VGG16 needed to be pre-trained on a large labeled image dataset. It can notably reduce training time and computational load.

\begin{figure*}[t!]
\centering
\includegraphics[scale=0.6]{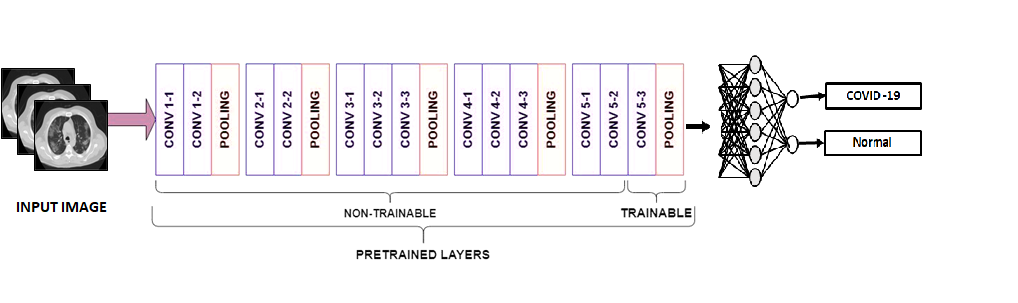}
\caption{VGG16 architecture for COVID-19 classification. (Image source: www.learndatasci.com)}
\label{fig:vgg16}
\end{figure*}

The architecture of VGG16 structure is shown in Figure~\ref{fig:vgg16}. The convolutional layer receives the input image of 224 × 224 and then propagates through a set of convolutional layers with a receptive field of 3 × 3. These are followed by a down sampling process and have five max-pooling layers through a set of specified convolutional layers and two fully connected layers with a fixed channel size ~\cite{arora2021transfer}. Every neuron is the fully connected layer that accepts the activations input from the preceding layer neuron. In VGG16, the average pooling layer AveragePooling2D layer, dropout layer, and fully connected layer are added ~\cite{halder2021covid}. Finally, an FC layer created a new set of classifications, namely COVID-19 and normal.

\section{Experiments and Results}
\label{sec:experimentsandresults}
\subsection{Implementation}

Performance metrics like Accuracy, AUC, F1-Score, Precision, and Recall are used to perform an experimental evaluation model. It is more helpful for evaluating the COVID-19 predictions. The formulae that calculate the performance metrics are shown below

$$Accuracy = \frac{TP + TN}{TP+FP+TN+FN}$$

We developed a public dataset, namely the All COVID Dataset, consisting of 14320 images grouped into 8535 COVID CT images and 5785 normal CT images. A whole dataset can be split into train, test, and validate to run the python programming code. This experimentation is done using the Kaggle data science environment freely. Both models are implemented using the python Keras library with TensorFlow as the backend. The models are evaluated four times with data split like 60\% training, 20\% testing, and 20\% validating with different epochs. The evaluated model behaviors are observed by accuracy.

\subsection{Results}
The experimental results of CNN and VGG 16 are shown in Figure 3, Figure 4 more helpful for automated COVID-19 identification on CT images. This study comprises 60\% training images and 20\% of testing, and 20\% validating images and setting the epoch value of 25 and 50 for CNN and VGG 16 models. Classifying the CT images for COVID identification achieved a prediction accuracy of 96.34\%, 96.99\% are shown in Table 2. The benefit of the proposed work is time requirements for CT image generation, and then models are trained and predicted, reducing the testing time more helpful to COVID-19 patients.

\begin{table}
\centering
\renewcommand{\arraystretch}{1.5}
\caption{Comparison of accuracy of CNN and VGG16 models.}
\begin{tabular}{p{2cm}p{8cm}p{2.5cm}}
\toprule
\textbf{Model} & \textbf{Dataset Size} & \textbf{Accuracy}  \\
\midrule
CNN & 14,320 (5,785 normal and 8,535 COVID-19) & 96.34\% \\
VGG16 & 14,320 (5,785 normal and 8,535 COVID-19) & 96.99\% \\
\bottomrule
\end{tabular}
\end{table}

\section{Conclusion}
\label{sec:conclusion}
Faster diagnosis of COVID-19 is the primary challenge for physicians today. COVID-19 patients can be quickly identified, which may contribute to addressing the current pandemic situations. For this purpose, we developed two deep learning models for identifying COVID-19 disease using chest CT images. We collected 14320 original COVID-19 CT images from various public databases and validated the performance of two deep learning models namely CNN and VGG16. We use these models to classify the CT image as COVID or normal. The accuracy of CNN and VGG16 models are 96.34\% and 97.40\%, respectively.
\bibliographystyle{splncs04}
\bibliography{references}
\end{document}